\documentclass{amsart}
\usepackage{amsmath,graphicx}

\begin{document}

\newcommand\re[1]{(\ref{#1})}
\newcommand{\pd}[2]{\frac{\partial #1}{\partial #2}}
\newcommand{\pdt}[3]{\frac{\partial^2 #1}{\partial #2 \partial #3}}
\newcommand{\mg}[1]{\partial_{#1}}
\newcommand{\F}[2]{F^{#1}_{\;#2}}
\renewcommand{\labelitemi}{--}

\title{Thermodynamic consistency of third grade finite strain elasticity}
\author{V\'an$^{1,2}$, P. and Papenfuss$^3$, C.}
\address{$^1$Department of Theoretical Physics\\
         KFKI, Research Institute of Particle and Nuclear Physics, Budapest,
         Hungary and \\
$^2$Department of Energy Engineering\\
        Budapest University of Technology and Economics, Hungary.\\
$^3$Technical University of Berlin, Sta\ss e des 17. Juni 135, 10623 Berlin, Germany; c.papenfuss@gmx.de.
}

\date{\today}

\begin{abstract}
   Thermodynamic framework of finite strain viscoelasticity with second order weak
    nonlocality in the deformation gradient is investigated. The application of Liu procedure leads to a class of
third grade elastic materials where  the second gradient of the stress appears
in the
elastic constitutive relation. Finally the dispersion relation of longitudinal plane waves is
    calculated in isotropic materials.
\end{abstract}

\maketitle

\section{Introduction}
Thermodynamic requirements are important in all theoretical approaches of
continuum mechanics.  The classical form of the Clausius-Duhem inequality
\cite{TruNol65b} does not allow higher than first grade elasticity  \cite{Gur65a} and 
plasticity \cite{AchSha95a} without any further ado. Therefore, new concepts 
- like higher order stresses, configurational forces - emerged to circumvent 
this condition and understand the thermodynamic compatibility of successful 
material models \cite{Gur00b,Mau99b,PapFor06a}.

In this paper we show that a weakly  nonlocal extension of the constitutive state space
does not contradict to the Second Law and leads to constitutive relations of higher grade finite strain elasticity
and viscoelasticity that are compatible with rigorous thermodynamic methods
and requirements. Our method is based on two basic observations, that are
different from the classical framework of Gurtin \cite{Gur65a}
\begin{itemize}
\item the entropy flux is a constitutive quantity,
\item for higher order weakly nonlocal state spaces the gradient of the balances
and other kinematic constraints result in further constraints on the entropy
inequality, 
 \end{itemize}

 The assumption of a \textit{constitutive entropy flux} is a straightforward
generalization of the Gibbs-Duhem inequality and one can prove that in simple cases leads to the classical 
form and to the classical results both in 
irreversible thermodynamics \cite{Van03a} or in thermoelasticity  
in particular \cite{Liu08a}. This generalization is well accepted and applied 
beyond mechanics \cite{JouAta92b,MulRug98b,Ver97b}. With the assumption of a 
constitutive entropy flux the Liu and Coleman-Noll procedures are equivalent 
(see \cite{TriAta08a} for a proof in a particular case).

It is also remarkable that further constraints in the Liu procedure result 
in more general constitutive functions \cite{MusEta08a}. Our main result,
the thermodynamic 
admissibility of the dependence of constitutive functions on space derivatives 
of the deformation gradient, is the consequence of this general property of 
the entropy inequality. Therefore we do not need to introduce higher order 
stresses (neither any other additional physical concepts) in advance, but at the 
end we will see, that some of the consequences of our method can be interpreted 
in that terms. The introduction  of the gradient of local constraints (e.g. 
balances) in non-equilibrium thermodynamics is a mathematical necessity in 
case of higher order weakly nonlocal constitutive state spaces - overlooked 
by Gurtin in \cite{Gur65a} - and has a unifying power to understand the role of the 
Second Law in several seemingly different theories of continuum physics (see
e.g. \cite{Van09a1} and the references therein).

In this paper we apply our method with Liu procedure in third grade elasticity
(\cite{TruNol65b}, p63), where the constitutive state space depends on the second space
derivative of the
deformation gradient. Therefore third grade elasticity is classified as
a second order weakly nonlocal theory. We prove the thermodynamic admissibility of a class of constitutive relations with two remarkable
properties
\begin{itemize}
\item the second order derivatives of the deformation appear without explicitly introducing double stress in
advance as an independent theoretical concept,
\item the second derivative of the stress is part of the nondissipative
stress-strain constitutive relation. This is similar to the suggestion of
Aifantis (see \cite{Aif03a} and the references therein).
\end{itemize}

We also derive a simple dispersion relation of longitudinal plane waves to demonstrate
the properties of the constitutive relation.

\section{Continuum in a Piola-Kirchhoff framework}

 All quantities are defined on the reference configuration. The substantial time
derivative is denoted by a dot and the material space derivative is \(\partial_i  \),
where \(i\in \{1,2,3\}\).  Higher order derivatives are denoted by more indices,
e.g. \(\partial_{ij}\) is the second gradient.  \(\chi^i\) is the motion, \(F^{i}_j=\partial_j\chi^i\) is the
deformation gradient.

According to the traditional concept of objectivity \cite{Nol58a} this  kinematic standpoint
ensures the objectivity of the whole treatment as long as the constitutive quantities 
depend on objective physical quantities. However,
the original mathematical formulation of the concept of objectivity by Noll is questionable
\cite{MatVan06a,MatVan07a} and a generalization based on precise spacetime
notions and a four-dimensional formalism was suggested to improve it
\cite{Ful08a3e,Van09m2}. In this work
 we do not apply this generalised objective framework, our results are derived
by usual three dimensional notions. However, we exploit some consequences of this generalization to simplify our calculation.
First of all it will be convenient to work in a Piola-Kirchhoff
framework, where the balances and the physical quantities are interpreted
in the reference configuration (see  \cite{Gur00b,Paw06a} for a similar treatment). The first Piola-Kirchhoff stress will
be denoted as a tensor. As we have mentioned above, a constitutive state space spanned by Noll-objective
physical quantities (e.g. right Cauchy-Green deformation) the objectivity of the whole treatment 
could be ensured.
However, it is more convenient to work with the deformation gradient, the material velocity and
the total energy as
basic variables. Moreover, it is not forbidden by our generalized notion of objectivity.
We will partially change to Noll-objective quantities at the end introducing the internal
energy with the usual definition and  show a particular stress-strain relation with
a  Cauchy deformation measure. 

Therefore in our treatment of the constitutive theory of third grade elastic
materials the constitutive state space is based on the following fields
\((v^i, \partial_j v^i,\) \(\partial_{jk} v^i, F^{i}_{\;j}, \partial_k F^{i}_{\;j},$
$\partial_{kl} F^{i}_{\;j},
e, \partial_i e)\). Here  \(v^{i}\) is the velocity, \(F^{i}_{\;j}\) is the deformation
gradient, and $e$ is the specific total energy. Our approach is second order
weakly nonlocal in the velocity and the deformation gradient, and first order
weakly nonlocal in the energy. It is usual to avoid introducing
velocity field by working with internal energy and internal energy balance
as an independent variable. However, the velocity and deformation gradient
fields form a single physical quantity therefore    we find instructive to show that the direct definition of internal
energy can be introduced at the end, and that the total energy filtered through Liu procedure can give
the same results when specified to local constitutive relations. This is
the approach that we have followed in case of relativistic fluids, where
Liu procedure was essential to distinguish  total and internal energies
\cite{Van08a}.

The well known kinematic relation between  the velocity and the deformation gradient \begin{equation}
\dot F^i_{\;j} - \partial_jv^i=0
\label{kincon}\end{equation}
is introduced as a constraint for the entropy inequality together with the balances.

The balance of linear momentum is 
\begin{equation}
        \rho_0\dot v^i - \partial_j T^{ij}=0^i.
\label{lmbal}\end{equation}
Here \(\rho_0\) is the material density and \(T^{ij} \) is the first
Piola-Kirchhoff stress, introduced as a tensor. The balance of total energy is
\begin{equation}
        \rho_0 \dot e+\partial_iq^i=0,
\label{ebal}\end{equation}

\noindent where $q^i$ is the energy flux. As we are working in a second order
weakly nonlocal constitutive state space the derivative of the kinematic relation \re{kincon}
and the momentum balance \re{lmbal} are additional
constraints according to the exploitation method of weakly nonlocal continuum
theories \cite{Van05a,Cim07a}:
\begin{eqnarray}
\mg{k}\F{i}{j} - \mg{kj}v^i &=& 0^{i}_{jk}, \label{dkincon}\\
\rho_0 \mg{j} \dot v^i  + \mg{jk}  T^{ik} &=& 0^i_{\;j}.\label{dmbal}
\end{eqnarray}
 
The gradient of the energy balance does not give
an additional constraint, because the constitutive state space is first order
weakly nonlocal in the energy. 

The entropy inequality requires that
\begin{equation}
        \rho_0 \dot s + \partial_i J^i\geq 0,
\label{sbal}\end{equation}
where $s$ is the specific entropy and $J^i$ is the material entropy flux.
Here we are looking for restrictions on the constitutive functions $T^{ij}, q^i, J^i$
in terms of the entropy density derivatives. It is important to see, that
the derivative of the momentum balance extends the process direction space,
which is spanned by the first and  also the second space derivatives of
the constitutive state space.

We introduce $\Lambda_i^j, \lambda_i, \kappa,
\Lambda_i^{jk}, \lambda^j_i$ Lagrange-Farkas
multipliers of the equations \re{kincon}-\re{dmbal} respectively.

Therefore the starting point of the Liu
procedure is the following inequality
\begin{gather}
0\leq\rho_0 \dot s +\partial_i J^i -
        \Lambda_i^j(\dot F^i_{\;j} - \partial_jv^i)-
        \lambda_i(\rho_0\dot v^i - \partial_jT^{ij})-
        \kappa(\rho_0 \dot e+\partial_iq^i)-\nonumber\\
        \Lambda_i^{jk}(\mg{k}\F{i}{j} - \mg{kj}v^i)-
        \lambda^j_i(\rho_0 \mg{j} \dot v^i  + \mg{kj}  T^{ik})=\nonumber\\
\rho_0\pd{s}{v^i}\dot v^i+
\rho_0\pd{s}{\mg{j}v^i}\mg{j}\dot v^i+
\rho_0\pd{s}{\mg{jk}v^i}\mg{jk}\dot v^i+
\rho_0\pd{s}{\F{i}{j}}\dot{\F{i}{j}}+
\rho_0\pd{s}{\mg{k}\F{i}{j}}\mg{k}\dot{\F{i}{j}}+
\nonumber\\
\rho_0\pd{s}{\mg{kl}\F{i}{j}}\mg{kl}\dot{\F{i}{j}}+
\rho_0\pd{s}{e}\dot e+
\rho_0\pd{s}{\mg{i}e}\mg{i} \dot e+
\nonumber\\
\pd{J^j}{v^{i}}\partial_j v^{i}+
\pd{J^k}{\mg{j}v^i}\mg{kj}v^i+
\pd{J^l}{\mg{kj}v^i}\mg{lkj} v^i+
\pd{J^k}{\F{i}{j}} \partial_k \F{i}{j}+
\pd{J^l}{\mg{k}\F{i}{j}}\mg{lk}{\F{i}{j}}+
\nonumber\\
\pd{J^m}{\mg{lk}\F{i}{j}}\mg{mlk}{\F{i}{j}}+
\pd{J^i}{e}\mg{i}e+
\pd{J^j}{\mg{i}e}\mg{ji}e -
\nonumber\\
\lambda\left(\rho_0\dot e +
\pd{q^j}{v^{i}}\mg{j} v^{i}+
\pd{q^k}{\mg{j}v^i}\mg{kj}v^i+
\pd{q^l}{\mg{kj}v^i}\mg{lkj} v^i+
\pd{q^k}{\F{i}{j}} \partial_k \F{i}{j}+
\pd{q^l}{\mg{k}\F{i}{j}}\mg{lk}{\F{i}{j}}+
\right.\nonumber\\\left.
\pd{q^m}{\mg{lk}\F{i}{j}}\mg{mlk}{\F{i}{j}}+
\pd{q^i}{e}\mg{i} e+
\pd{q^j}{\mg{i}e}\mg{ji} e\right) -
\nonumber\end{gather}
\begin{gather}
\lambda_r\left(\rho_0\dot v^i-
\pd{T^{rj}}{v^{i}}\mg{j} v^{i}-
\pd{T^{rk}}{\mg{j}v^i}\mg{kj}v^i-
\pd{T^{rl}}{\mg{kj}v^i}\mg{lkj} v^i-
\pd{T^{rk}}{\F{i}{j}} \partial_k \F{i}{j}-
\pd{T^{rl}}{\mg{k}\F{i}{j}}\mg{lk}{\F{i}{j}}-
\right.\nonumber\\\left.
\pd{T^{rm}}{\mg{lk}\F{i}{j}}\mg{mlk}{\F{i}{j}}-
\pd{T^{ri}}{e}\mg{i}e
-\pd{T^{rj}}{\mg{i}e}\mg{ji}e
\right) -
\nonumber\\
\Lambda_i^j(\dot F^i_{\;j} - \partial_jv^i)-
\nonumber\\
\lambda^s_r\left(\rho_0\mg{s} \dot{v}^r-
\pd{T^{rj}}{v^{i}}\mg{sj} v^{i}-
\mg{j} v^{i}\mg{s}\left[\pd{T^{rj}}{v^{i}}\right]-
\pd{T^{rk}}{\mg{j}v^i}\mg{skj}v^i-
\mg{kj}v^i \mg{s}\left[\pd{T^{rk}}{\mg{j}v^i}\right]-
\right.\nonumber\\\left.
\pd{T^{rl}}{\mg{kj}v^i}\mg{slkj} v^i-
\mg{lkj}v^i \mg{s}\left[\pd{T^{rl}}{\mg{kj}v^i}\right]-
\pd{T^{rk}}{\F{i}{j}} \mg{sk} \F{i}{j}-
\mg{k} \F{i}{j} \mg{s}\left[\pd{T^{rk}}{\F{i}{j}}\right]-
\right.\nonumber\\\left.
\pd{T^{rl}}{\mg{k}\F{i}{j}}\mg{slk}{\F{i}{j}}-
\mg{lk}{\F{i}{j}} \mg{s}\left[\pd{T^{rl}}{\mg{k}\F{i}{j}}\right]-
\pd{T^{rm}}{\mg{lk}\F{i}{j}}\mg{smlk}{\F{i}{j}}-
\mg{mlk}{\F{i}{j}} \mg{s}\left[\pd{T^{rm}}{\mg{lk}\F{i}{j}}\right]-
\right.\nonumber\\\left.
\pd{T^{ri}}{e}\mg{si}e-
\mg{i}e \;\mg{s}\left[\pd{T^{ri}}{e}\right]
-\pd{T^{rj}}{\mg{i}e}\mg{sji}e -
\mg{ji}e \;\mg{s}\left[\pd{T^{rj}}{\mg{i}e}\right]
\right) -
\nonumber\\
\Lambda_i^{jk}(\mg{k}\dot F^i_{\;j} - \partial_{kj}v^i)
\label{bineq}\end{gather}

The time derivative related Liu equations imply that Lagrange-Farkas multipliers are 
determined by the entropy derivatives,
and that the specific entropy and the stress do not depend on the highest derivatives
of the constitutive state space $s=s(v^i, \partial_j v^i,F^{i}_{\;j}, \partial_k F^{i}_{\;j},
e)$.

Then, the Liu equations related to the highest order space derivatives   determine the entropy flux in the following form
\begin{equation}
J^i = \pd{s}{e}q^i -
     \pd{s}{\mg{i}v^r}\left(\pd{T^{jr}}{e}\mg{j}e +
        \pd{T^{kr}}{\mg{l}v^m}\mg{lk}v^{m}+
        \pd{T^{kr}}{\mg{l}\F{n}{m}}\mg{lk}\F{n}{m}\right)+
        K^{i}.
\label{sf}\end{equation}
Here $K^i=K^i(v^i, \partial_j  v^i,F^{i}_{\;j}, \partial_k F^{i}_{\;j}, e)$ is the
extra entropy flux with the above denoted restricted functional dependencies.
Then we introduce several convenient assumptions in order to get a simple
and solvable form of the dissipation inequality. First of all we define the
internal energy of the third grade viscoleastic material including an isotropic kinetic energy contribution of  deformation gradient rate
and also the classical heat flux \(\hat
q\):
\begin{equation}
u := e- \frac{1}{2}v^iv_i - \frac{\alpha_1}{2} (\dot{\F{i}{i}})^2 -
     \frac{\alpha_2}{2}\dot{\F{i}{j}}\dot{\F{j}{i}}, \qquad 
\qquad      \hat
q^i :=q^i+v_jT^{ij}.
\end{equation}
 Morever, we may observe that a particular choice of the extra
entropy flux  reduces the dissipation inequality to a solvable form. Therefore we
assume that
$$
K^i :=\pd{s}{e}v_jT^{ij}-\pd{s}{\mg{i}\F{k}{j}}\mg{j}v^k.
$$

Finally the temperature \(\theta\) is defined by the entropy derivatives  
$\pd{s}{e}=\pd{s}{u}=\frac{1}{\theta}$
and  the free energy as $\psi(\mathbf{F},\nabla\mathbf{F}):=
u -\theta s$. Then we obtain   
\begin{gather}
\theta \sigma_S =
 \theta \hat q^i\partial_i \frac{1}{\theta} +
 \mg{j}v^{i}\left(
        T_i^{j} -
        \pd{\psi}{\F{i}{j}}  -
        \alpha_1\mg{lk}T^{lk}\delta^j_i -
        \alpha_2 \partial^j_{\;\;k}T^{k}_{\;\;i} +
        \mg{k}\pd{\psi}{\mg{k}\F{i}{j}} \right) \geq 0.
\label{dineq1}\end{gather}

The first two terms in the parenthesis of the above expression are the classical
terms from second grade elasticity. The very last term in \re{dineq1} resembles
the double stress relation that one can get by virtual power techniques.

In the  non dissipative case assuming constant material parameters we get
a constitutive relation of third grade
elasticity in the following form
\begin{equation}
\mathbf T -
        \nabla\cdot\left(\alpha_1(\nabla\cdot\mathbf T)\mathbf I  +
        \alpha_2 \nabla\mathbf T \right)=
         \pd{\psi}{\mathbf F}- \nabla\cdot
        \pd{\psi}{\nabla\mathbf F}.
\label{ssrel}\end{equation}

\section{Simple waves}

We may calculate the dispersion relation of a one dimensional plane wave
in the small strain approximation  in order to check some consequences of the above stress-strain relation.

Let us assume a usual isotropic quadratic free energy for symmetric strains \(\epsilon^i_j
= \frac{1}{2}(\F{i}{j}+\F{j}{i} - 2 \delta^i_j)\), that is second order isotropic
quadratic also in the gradient of the symmetric strains. Then according to
representation theorems only
two additional material parameters $a_1,a_2$ appear  in the free energy function
\cite{Min65a}
\begin{equation}
\psi(\epsilon^i_j, \mg{k}\epsilon^i_j) =
        \frac{\lambda}{2} (\epsilon^i_i)^2+
        \mu \epsilon^i_j\epsilon^i_j +
        \frac{a_1}{2} \mg{i}\epsilon^j_j\partial^i\epsilon^k_k+
        \frac{a_2}{2} \mg{i}\epsilon^j_k\partial^i\epsilon^k_j.
\end{equation}

Then we get
\begin{equation}
T_i^{j} -
        \alpha_1\mg{k}\mg{l}T^{lk}\delta^j_i -
        \alpha_2 \mg{k}\partial^jT^{k}_{\;\;i} =
        \lambda \epsilon^k_k \delta^j_i +
        2\mu \epsilon^j_i -
        a_1 (\partial_{k}^k \epsilon^l_l)\delta^j_i -
        a_2 \partial_k^k(\epsilon^j_i).\end{equation}

Let us investigate the simplest one dimensional case and reduce the treatment
to one component of the above tensorial equation. Introducing the notation
$T=T^{11}$ and $\epsilon=\epsilon^{11}$ and also denoting $\partial_1$ by
a dash we get
$$
T-\alpha T'' =  \hat\lambda\epsilon -a\epsilon''.
$$
where  $\alpha=\alpha_1+\alpha_2$, $\hat\lambda=\lambda+2\mu$
and $a=a_1+a_2$.
This constitutive relation is coupled to the balance of momentum that is
in our case
$$
\rho_0 \ddot \epsilon - T''=0.
$$

Assuming constant material parameters we get the following dispersion relation
$$
\omega^2 = \frac{k^2(\hat\lambda + a k^2)}{\rho_0(1+\alpha k^2)}.
$$

\begin{figure}
\includegraphics[width=0.8\textwidth]{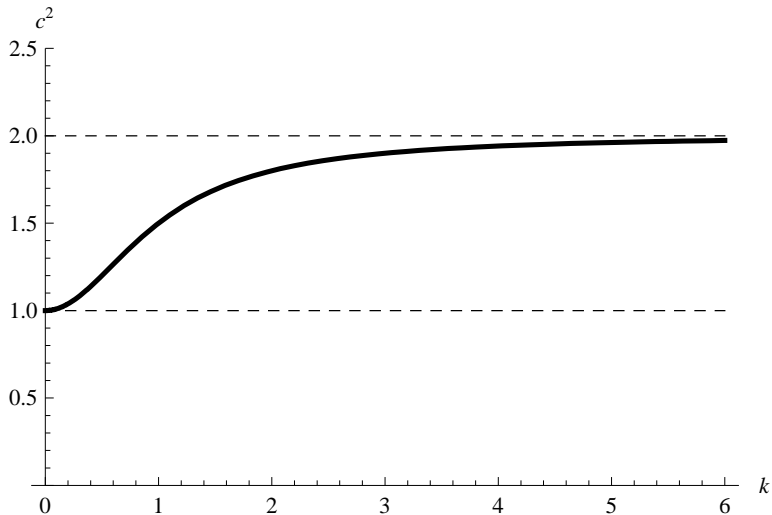}
\caption{\label{Fig1}
The square of the phase velocity $v^2 = (\omega(k)/k)^2 $  with the parameter
values $\rho_0 =1$ and $\hat \lambda=1$, $\alpha=1$ and $a=2$.}
\end{figure}

The characteristic feature of this dispersion relation is that the small
wavelength and large wavelength limits result in  finite acoustic phase
velocities: \(\\ \lim_{k\rightarrow 0} \frac{\omega(k)}{k} = \sqrt{\hat\lambda/\rho_0}\)
and \(\lim_{k\rightarrow \infty}\frac{\omega(k)}{k} = \sqrt{a/(\alpha\rho_0)}\)
as it is demonstrated on Figure \ref{Fig1}. This kind of behaviour
is  a  property of the double wave equation \cite{PorAta09a}.
Double wave equations are introduced by microstructural considerations
e.g. in microstrain theories or internal variable theories \cite{Min65a,BerEta10a}.

\section{Summary}

In this paper we have investigated a weakly nonlocal extension of viscoelasticity
up to second order in the deformation gradient. The entropy flux was considered as a constitutive
quantity and we applied Liu procedure introducing also the gradients of the
balance of momentum and of the kinematic relation (\ref{kincon}) as constraints of the entropy
balance.

The calculations were performed in the Piola-Kirchhoff framework. The constitutive state space was chosen
according to the generalization of the Noll principle of frame indifference
introducing the components of the velocity-deformation gradient mixed four-tensor
and their first and second space derivatives as constitutive variables. 

 We have given a complete solution of the Liu equations and obtained a particular form of the entropy flux and the dissipation
inequality. In order to solve the  dissipation inequality we have
introduced some simplifications. In particular a quadratic form of the kinetic
energy and some further related rearrangements resulted in a form where Onsagerian
fluxes and forces can be identified. The constitutive relation of the non-dissipative mechanical material contains the gradients of the pressure in addition to
the classical terms and the usual form of the constitutive relation of the double stress appeared without postulating
such term in advance.

It is important to note that the  extension of the constitutive state space toward
higher order gradients does not change the obtained form of \re{dineq1} as
long as only the first derivatives of  \re{kincon} and \re{lmbal} are introduced
as additional constraints. This extension results in simple explicit solutions of
the  non-dissipative differential stress-deformation relation and then \re{ssrel}
can be considered as a constitutive relation of third grade elasticity. 

We have calculated also a one dimensional dispersion relation and concluded,
that it is similar to the dispersion relation of some microstructured materials
as one could expect in case of higher grade solids (see e.g. \cite{ForSie06a}).

As we have mentioned in the introduction,  stress-strain relations similar
to \re{ssrel} were
already proposed in the literature \cite{Aif03a}. However, there the motivation
was to remove stress singularities with a kind of ad hoc "reaction diffusion" form. Here we have shown that this kind of extension
is compatible with a weakly nonlocal thermodynamic framework, there are natural
boundary conditions coming from the requirement of vanishing entropy flux
and a physical origin of stress derivative terms is the modified (isotropic) kinetic
energy, related to the rate of the deformation gradient.

\section{Acknowledgement}

The authors thank  Arkadi Berezovski for the fruitful discussions.
The work was supported by the grant Otka K81161.

\bibliographystyle{unsrt}

\begin{thebibliography}{10}

\bibitem{TruNol65b}
C.~Truesdell and W.~Noll.
\newblock {\em The Non-Linear Field Theories of Mechanics}.
\newblock Springer Verlag, Berlin-Heidelberg-New York, 1965.
\newblock Handbuch der Physik, III/3.

\bibitem{Gur65a}
M.~E. Gurtin.
\newblock Thermodynamics and the possibility of spatial interaction in elastic
  materials.
\newblock {\em Archive for Rational Mechanics and Analysis}, 19:339--352, 1965.

\bibitem{AchSha95a}
A.~Acharya and T.~G. Shawki.
\newblock Thermodynamic restrictions on constitutive equations for
  second-deformation-gradient inelastic behaviour.
\newblock {\em Journal of Mechanics and Physics of Solids}, 43:1751--1772,
  1995.

\bibitem{Gur00b}
M.~E. Gurtin.
\newblock {\em Configurational forces as basic concepts of continuum physics}.
\newblock Springer, New York-etc., 2000.

\bibitem{Mau99b}
G.~Maugin.
\newblock {\em The thermomechanics of nonlinear irreversible behaviors ({A}n
  introduction)}.
\newblock World Scientific, Singapure-New Jersey-London-Hong Kong, 1999.

\bibitem{PapFor06a}
C.~Papenfuss and S.~Forest.
\newblock Thermodynamical frameworks for higher grade material theories with
  internal variables or additional degrees of freedom.
\newblock {\em Journal of Non-Equilibrium Thermodynamics}, 31(4):319--353,
  2006.

\bibitem{Van03a}
P.~V\'an.
\newblock Weakly nonlocal irreversible thermodynamics.
\newblock {\em Annalen der Physik (Leipzig)}, 12(3):146--173, 2003.
\newblock (cond-mat/0112214).

\bibitem{Liu08a}
I-Shih Liu.
\newblock Entropy flux relation for viscoelastic bodies.
\newblock {\em Journal of Elasticity}, 90(3):259--270, 2008.

\bibitem{JouAta92b}
D.~Jou, J.~Casas-V\'azquez, and G.~Lebon.
\newblock {\em Extended Irreversible Thermodynamics}.
\newblock Springer Verlag, Berlin-etc., 1992.
\newblock 3rd, revised edition, 2001.

\bibitem{MulRug98b}
I.~M\"uller and T.~Ruggeri.
\newblock {\em Rational Extended Thermodynamics}, volume~37 of {\em Springer
  Tracts in Natural Philosophy}.
\newblock Springer Verlag, New York-etc., 2nd edition, 1998.

\bibitem{Ver97b}
J.~Verh\'as.
\newblock {\em Thermodynamics and {R}heology}.
\newblock Akad\'emiai Kiad\'o and Kluwer Academic Publisher, Budapest, 1997.

\bibitem{TriAta08a}
V.~Triani, C.~Papenfuss, V.~A. Cimmelli, and W.~Muschik.
\newblock Exploitation of the {S}econd {L}aw: {C}oleman-{N}oll and {L}iu
  procedure in comparison.
\newblock {\em Journal of Non-Equilibrium Thermodynamics}, 33:47--60, 2008.

\bibitem{MusEta08a}
W.~Muschik, Vita Triani, and Christina Papenfuss.
\newblock Exploitation of the dissipation inequality, if some balances are
  missing.
\newblock {\em Journal of Mechanics of Materials and Structures},
  3(6):1125--1133, 2008.

\bibitem{Van09a1}
V\'an P.
\newblock Weakly nonlocal non-equilibrium thermodynamics - variational
  principles and {S}econd {L}aw.
\newblock In Ewald Quak and Tarmo Soomere, editors, {\em Applied Wave
  Mathematics (Selected Topics in Solids, Fluids, and Mathematical Methods)},
  chapter III, pages 153--186. Springer-Verlag, Berlin-Heidelberg, 2009.
\newblock arXiv:0902.3261.

\bibitem{Aif03a}
E.~C. Aifantis.
\newblock Update on a class of gradient theories.
\newblock {\em Mechanics of Materials}, 35:259--280, 2003.

\bibitem{Nol58a}
W.~Noll.
\newblock A mathematical theory of the mechanical behavior of continuous media.
\newblock {\em Archives of Rational Mechanics and Analysis}, 2:197--226,
  1958/59.

\bibitem{MatVan06a}
T.~Matolcsi and P.~V\'an.
\newblock Can material time derivative be objective?
\newblock {\em Physics Letters A}, 353:109--112, 2006.
\newblock math-ph/0510037.

\bibitem{MatVan07a}
T.~Matolcsi and P.~V\'an.
\newblock Absolute time derivatives.
\newblock {\em Journal of Mathematical Physics}, 48:053507--19, 2007.
\newblock math-ph/0608065.

\bibitem{Ful08a3e}
T.~F\"ul\"op.
\newblock A new interpretation of the kinematics of continua.
\newblock In T.~F\"ul\"op, editor, {\em New results in continuum physics},
  volume~8 of {\em Notes on Engineering Geology and Rock Mechanics}, chapter~3,
  pages 55--99. BME Publisher, Budapest, 2008.
\newblock in Hungarian.

\bibitem{Van09m2}
P.~V\'an.
\newblock Four dimensional treatment of field quantities in non-relativistic
  spacetime.
\newblock 2009.

\bibitem{Paw06a}
I.~Paw\l~ow.
\newblock Thermodynamically consistent {C}ahn-{H}illiard and {A}llen-{C}ahn
  models in elastic solids.
\newblock {\em Discrete and Continuous Dynamical Systems}, 15(4):1169--1191,
  2006.

\bibitem{Van08a}
P.~V\'an.
\newblock Internal energy in dissipative relativistic fluids.
\newblock {\em Journal of Mechanics of Materials and Structures},
  3(6):1161--1169, 2008.
\newblock Lecture held at TRECOP'07, arXiv:07121437 [nucl-th].

\bibitem{Van05a}
P.~V\'an.
\newblock Exploiting the {S}econd {L}aw in weakly nonlocal continuum physics.
\newblock {\em Periodica Polytechnica, Ser. Mechanical Engineering},
  49(1):79--94, 2005.
\newblock (cond-mat/0210402/ver3).

\bibitem{Cim07a}
V.~A. Cimmelli.
\newblock An extension of {L}iu procedure in weakly nonlocal thermodynamics.
\newblock {\em Journal of Mathematical Physics}, 48:113510, 2007.

\bibitem{Min65a}
R.~D. Mindlin.
\newblock Second gradient of strain and surface-tension in linear elasticity.
\newblock {\em International Journal of Solids and Structures}, 1:417--438,
  1965.

\bibitem{PorAta09a}
A.~V. Porubov, E.~L. Aero, and G.~A. Maugin.
\newblock Two approaches to study essential nonlinear and dispersive properties
  of the internal structure of materials.
\newblock {\em Physical Review E}, (79):046608, 2009.

\bibitem{BerEta10a}
A.~Berezovski, J.~Engelbrecht, and G.~A. Maugin.
\newblock Generalized thermomechanics with dual internal variables.
\newblock {\em Archive of Applied Mechanics}, 2010.
\newblock online first.

\bibitem{ForSie06a}
S.~Forest and R.~Sievert.
\newblock Nonlinear microstrain theories.
\newblock {\em International Journal of Solids and Structures}, 43:7224--7245,
  2006.

\end{thebibliography}

\end{document}